\documentclass[conference,a4paper]{APSIPA2021}

\usepackage{amsmath}
\usepackage{graphicx}
\usepackage{multirow}
\usepackage{threeparttable}
\usepackage{hyperref}
\usepackage{cleveref}
\usepackage{optidef}
\usepackage{amsfonts}
\usepackage{cite}

\crefname{figure}{Fig.}{Figs.}

\usepackage{siunitx}
\DeclareSIUnit \decibelA {dB(A)}
\DeclareSIUnit \decibelC {dB(C)}

\usepackage{geometry}
\usepackage{fancyhdr}

\fancypagestyle{firststyle}{
  \fancyhf{}
  \fancyhead[C]{2024 Asia Pacific Signal and Information Processing Association Annual Summit and Conference (APSIPA ASC)}
}

\begin{document}

\title{A Real-Time Platform for Portable and Scalable Active Noise Mitigation for Construction Machinery}

\author{
\authorblockN{
Woon-Seng Gan\authorrefmark{1}, 
Santi Peksi\authorrefmark{2},
Chung Kwan Lai\authorrefmark{3},
Yen Theng Lee\authorrefmark{4},
Dongyuan Shi\authorrefmark{5} and
Bhan Lam\authorrefmark{6}
}

\authorblockA{
\authorrefmark{1}\authorrefmark{2}\authorrefmark{4}\authorrefmark{5}\authorrefmark{6}
School of Electrical and Electronic Engineering, Nanyang Technological University, Singapore\\
E-mail: \{ewsgan\authorrefmark{1}, speksi\authorrefmark{2}, dongyuan.shi\authorrefmark{4}, bhan lam\authorrefmark{5}\} @ntu.edu,sg}

\authorblockA{
\authorrefmark{3}
e190091@e.ntu.edu.sg}

\authorblockA{
\authorrefmark{3} Institute of Sound and Vibration Research, University of Southampton, UK\\
C.K.Lai@soton.ac.uk\authorrefmark{3}
}
}

\maketitle
\thispagestyle{firststyle}
\pagestyle{fancy}

\begin{abstract}
 This paper introduces a novel portable and scalable Active Noise Mitigation (PSANM) system designed to reduce low-frequency noise from construction machinery. The PSANM system consists of portable units with autonomous capabilities, optimized for stable performance within a specific power range. An adaptive control algorithm with a variable penalty factor prevents the adaptive filter from over-driving the anti-noise actuators, avoiding non-linear operation and instability. This feature ensures the PSANM system can autonomously control noise at its source, allowing for continuous operation without human intervention. Additionally, the system includes a web server for remote management and is equipped with weather-resistant sensors and actuators, enhancing its usability in outdoor conditions. Laboratory and in-situ experiments demonstrate the PSANM system's effectiveness in reducing construction-related low-frequency noise on a global scale. To further expand the noise reduction zone, additional PSANM units can be strategically positioned in front of noise sources, enhancing the system's scalability.The PSANM system also provides a valuable prototyping platform for developing adaptive algorithms prior to deployment. Unlike many studies that rely solely on simulation results under ideal conditions, this paper offers a holistic evaluation of the effectiveness of applying active noise control techniques directly at the noise source, demonstrating realistic and perceptible noise reduction. This work supports sustainable urban development by offering innovative noise management solutions for the construction industry, contributing to a quieter and more livable urban environment.
\end{abstract}
\vspace{1em} 
\begin{keywords}
 Active Noise Control(ANC), construction machinery, output power constraint, in-situ noise control, modified filtered reference least mean square algorithm
\end{keywords}

\section{Introduction}

The rapid evolution of modern construction methodologies and machinery has significantly expedited urban development and infrastructure projects worldwide. However, this progress incurs an environmental cost, particularly in the form of noise pollution \cite{Murphy2014}. Construction machinery, including generators, exhaust systems, and transformers, are primary contributors to urban noise, adversely affecting public health and the environment. The World Health Organization (WHO) identifies noise pollution as the second-largest environmental cause of health issues, following air quality concerns  \cite{WHO2020}.

Active Noise Mitigation (ANM) technologies complement traditional noise control strategies, providing greater effectiveness for low-frequency noise, while passive methods, such as noise barriers, excel at higher frequencies. Unlike passive methods, ANM systems utilize electronic means to generate a sound wave that matches the original noise in amplitude but is opposite in phase, effectively canceling it out through destructive interference. This principle, known as active noise control (or mitigation), has been extensively researched and developed over the past few decades, finding applications across various industries \cite{kajikawa2012recent,lam2021ten,shen2024principle,shi2023active,shi2016open,chang2023complete,shen2023study}. One of the earliest uses of active noise control technology was in aerospace, where it helped reduce cabin noise in commercial aircraft \cite{Mylonas2024}. The technology has since been adapted for the automotive industry, with manufacturers integrating ANM systems to enhance driving experiences by reducing engine and road noise \cite{Oh2018}.

In construction, ANM technologies offer promising solutions for managing the complex noise generated by machinery. These systems can be integrated into machinery or deployed as mobile and scalable units, enhancing flexibility in noise management strategies. A pilot study demonstrated significant noise level reductions at a construction site using ANM systems, indicating potential for broader application in the industry  \cite{Mostafavi2023}. However, it is important to note that these experiments have primarily been conducted in controlled laboratory settings, using loudspeakers to replicate engine or machinery noise, and have yet to be tested under real on-site conditions.

The portable and scalable Active Noise Mitigation (PSANM) presents substantial noise reduction advantages over traditional passive noise control methods, especially for low-frequency noise, which is challenging to mitigate effectively using conventional means. A comprehensive review outlines the effectiveness of various passive noise control materials and designs, noting their widespread application in urban construction settings \cite{Hansen2021}. However, while passive methods effectively address high-frequency noise, they often fall short against low-frequency noise generated by construction machinery \cite{MostafaMir2022,Sohrabi2020}. Additionally, physical constraints and aesthetic considerations in urban environments limit the applicability of extensive passive noise mitigation structures. The European Environment Agency emphasizes that while passive measures are crucial for noise management, they cannot singularly resolve the multifaceted challenges of urban noise pollution, underscoring the need for innovative solutions \cite{EEA2023}.

This paper offers an overview of the current noise pollution landscape caused by construction activities and introduces the fundamental principles of PSANM technology. It examines PSANM's potential in controlling low-frequency construction noise without the need of a passive element. Furthermore, the paper delves into the technical aspects of the developed an adaptive active noise control system, which has been tested in both laboratory and on-site environments with noise generators. It also explores future directions for innovation and wider adoption of PSANM systems.

By addressing the critical issue of noise pollution in construction, this paper contributes to the ongoing discussion surrounding sustainable urban development and the role of technology in mitigating environmental impacts. This paper is organized as follows: In \Cref{sec:architecture}, a real-time PSANM system architecture is being presented and follows by detailed description on the hardware components, adaptive signal processing algorithms, software elements, and system integration with remote control of the PSANM. \Cref{sec:lab} and \labelcref{sec:insitu} outline our controlled laboratory and in-situ experiments, respectively, and highlight the noise reduction limit and its spatial zone. We conclude our research work in \Cref{sec:conclusion}.

\section{Proposed Real-Time PSANM} \label{sec:architecture}
\begin{figure}[t]
\begin{center}
\includegraphics[width=90mm]{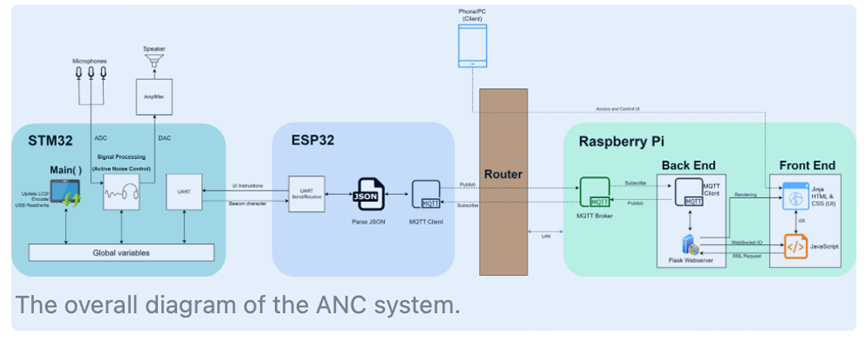}
\end{center}
\caption{The overall block diagram of the PSANM System, which consists of the active noise mitigation processing module, communication module for remote operational mode control, and module that performs data monitoring and performance evaluation}
\label{fig:overall}
\vspace*{-30pt}
\end{figure}

To develop a portable and scalable Active Noise Mitigation (PSANM) system aimed at reducing low-frequency noise emissions from construction machinery, we designed a series of portable units powered by embedded systems, including the STM32 microcontroller, ESP32 processing module, and Raspberry Pi 4 (RPi4) \cite{raspberrypi2024}, as shown in \Cref{fig:overall}. The STM32 microcontroller is programmed with adaptive algorithm that is optimized to operate within a specified power output range, ensuring both distortion-free performance and system stability \cite{stm32h72024}. In order to achieve remote access and monitoring of the PSANM units, we have programmed the ESP32 module as a WiFi module to connect to a router, which is linked to the RPi4 via LAN connection \cite{espressif2024}. The RPi4 serves dual functions, hosting both the MQTT broker for efficient message handling and the Web Server. The PSANM system can adapt to various types of machinery noise, effectively managing frequencies below \SI{200}{\hertz} with noise levels exceeding \SI{90}{\decibelC}.

\subsection{Key Features and Innovations}

One of the standout features of the PSANM system is its integrated web server, which enables streamlined remote management and monitoring. This web server provides significant operational flexibility, allowing users to access and control the system from any location with an internet connection. The remote access capability is crucial for managing PSANM units in large or hard-to-reach construction sites, ensuring continuous and efficient noise control operations.

To ensure durability and reliability in challenging outdoor environments, the PSANM units are housed in weather-resistant enclosures. These enclosures protect the sensitive electronic components from adverse weather conditions such as rain, dust, and extreme temperatures, ensuring the longevity and consistent performance of the system, as illustrated in Fig. 2. A photo of the dual PSANM units is shown in Fig. 3.

The system's active noise mitigation module uses reference and error microphones, which feed into an STM32 microcontroller equipped with a Nucleo ARM Cortex-M7 processor. The signals from these microphones are amplified and interfaced with the analog-to-digital converters (ADC) on the STM32 development board. The digitized signals are processed by a set of real-time adaptive algorithms, which will be further detailed in Section II.B. The processed anti-noise signals are then delivered to a secondary source, a 10-inch, 400 Watt (RMS) subwoofer with a sensitivity of 85.7 dB 1W/1m.

 To perform the initialization stage of secondary path calibration before the actual production of the anti-noise signal, secondary path estimation must be carried out in the noisy environment. Given the primarily tonal nature of the machinery noise, an uncorrelated and broadband training signal is used for this estimation. Experiments have shown that in-situ secondary path estimation is feasible and does not result in significant degradation of noise reduction performance. The following sections will further elaborate on the enhanced adaptive algorithm that can be applied in a real-world environment.

\subsection{Adaptive Algorithms}
The conventional Filtered reference Least Mean Square (FxLMS) Algorithm \cite{Hansen2021,Kuo1999,shi2020algorithms,luo2022hybrid,shi2019practicalA,yang2020stochastic,shi2016systolic}, remains the most effective and practical adaptive algorithm for real-time implementation. The PSANM system adopts the baseline FxLMS algorithm for both feedforward and feedback modes to better address different noise characteristics.

In the feedforward mode, a shared reference microphone is required, and this reference information must be transmitted to all PSANM units either by wire or wirelessly. Additionally, cross-channel secondary path information is necessary to achieve centralized feedforward control among the PSANM units. In contrast, the feedback mode allows each PSANM unit to be self-sufficient with its own reference and error signals, enabling decentralized noise control. The feedback mode is particularly suited for handling narrow-band construction noise and offers better scalability and flexibility in different noise regions compared to the feedforward mode, which can handle a broader frequency range but at a higher implementation cost.

Currently, there is a growing research interest in distributed active noise control to expand the spatial region of noise control. The PSANM system can play a crucial role in experimenting with the real-time performance of distributed noise control, contributing significantly to this emerging field.

\subsection{Adaptive Algorithm with Variable Penalty Factor}
In an PSANM system, the disturbance $d(n)$ can be suppressed by the anti-noise $y'(n)$ from the control signal $y(n)$ passed through the secondary path $s(n)$. The control signal is the result of processing the reference signal $x(n)$ by using the control filter $\mathbf{w}(n)$. To circumvent the output-saturation issue of practical PSANC systems~\cite{shi2017effect,shi2019practical,guo2024survey}, restricting the variance of the control signal has been proven to be an efficient and feasible solution. Hence, the objective of PSANM algorithm can be represented as a quadratically constrained quadratic program (QCQP) that minimizes the mean squared error signal under the limited output power~\cite{shi2019optimal,shi2021optimal,shi2019two,shi2021optimalA,gan2023practical}:

\begin{mini!}|l| 
	  {\mathbf{w}}{ f(\mathbf{w}) =\mathbb{E}\left[e^2(n)\right],}{}{}\label{optimization_a}
	  \addConstraint{g(\mathbf{w})=\mathbb{E}\left[y^2(n)\right]}{\le\rho^2.}{}\label{optimization_b}
\end{mini!}

Figure 4 presents the block diagram for implementing a variant of the feedforward FxLMS algorithm, known as the Minimum Output Variance FxLMS (MOV-FxLMS) algorithm \cite{Lai2023}, which utilizes an optimal time-varying penalty factor to prevent output saturation in the PSANM system. Additionally, a feedback version of the FxLMS algorithm, based on internal model control, has been programmed as an alternative option for the PSANM system.

A real-time penalty factor, $\alpha(n)$, is computed for each time frame of M samples, based on the varying power gain of the secondary path, $G_s(n)$, and the power of the estimated disturbance, $d(n)$, as shown in the following
\begin{equation}\label{eq25:TVPenalty}
    \alpha(n) = \text{max}\Biggl\{ G_s(n)\Biggl(\sqrt{\frac{\sum_{m=0}^{M-1}{\hat{d}}^2(n-m)}{MG_s(n)\rho^2}}-1 \Biggl) , 0\Biggl\},
\end{equation}
where 0 is the minimal value to avoid negative $\alpha(n)$. Therefore, the improved optimal time-varying penalty factor MOV-FxLMS algorithm can be rewritten as
\begin{equation}\label{eq26:VPFmovFxLMS}
    \mathbf{w}(n+1) = \mathbf{w}(n) +\mu[e(n)\mathbf{x'}(n) - \alpha(n) y(n)\mathbf{x}(n)].
\end{equation}

The advantage of using a real-time penalty factor is that it helps the MOV-FxLMS algorithm achieve optimal control under output constraints, even with abruptly changing noise. This capability is critical for the autonomous operation of the PSANM system.

\begin{figure}[t]
\begin{center}
\includegraphics[width=40mm]{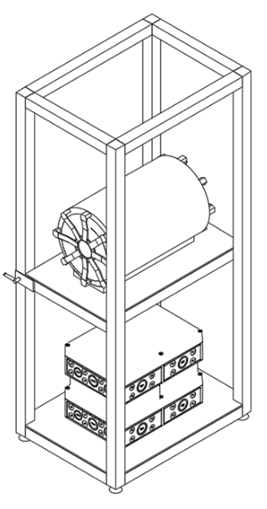}
\label{fig:fig2}
\end{center}
\caption{Isometric view of a single PSANM device }
\vspace*{-3pt}
\end{figure}

\begin{figure}[t]
\begin{center}
\includegraphics[width=85mm]{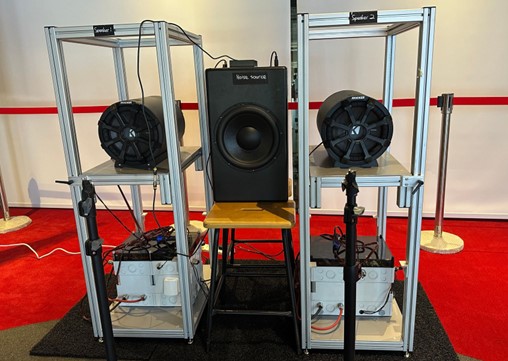}
\end{center}
\label{fig:fig3}
\caption{The demo setup of dual-PSAMN device that place in front of the noise source (at center)}
\vspace*{-3pt}
\end{figure}

\begin{figure}[t]
\begin{center}
\includegraphics[width=90mm]{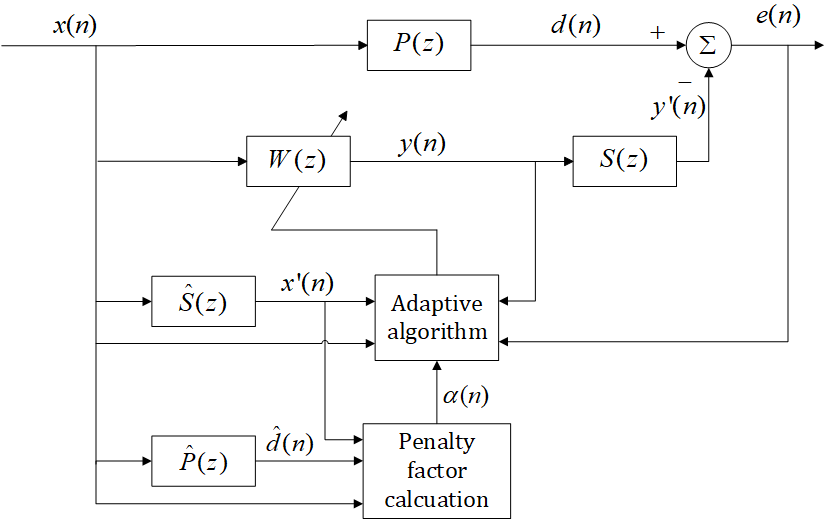}
\end{center}
\caption{The MOV-FXLMS (feedforward) algorithm that is being programmed into the PSANM system to prevent output saturation and allow autonomous operation}
\vspace*{-3pt}
\end{figure}

\subsection{System Integration and Remote Control}
Remote control capabilities further enhance the system's flexibility and user convenience. We employ an ESP32 module as a Wi-Fi bridge, facilitating communication between the PSANM system and the web server via the MQTT (Message Queuing Telemetry Transport) protocol for network communications. This setup allows the Wi-Fi module to connect to a router, which is directly linked to the Raspberry Pi via LAN (Local Area Network). The ESP32 module updates the STM32 with any system status updates via serial communication through UART (Universal Asynchronous Receiver-Transmitter) ports, once the updates are received from the MQTT broker. The Raspberry Pi hosts both the MQTT broker for efficient message handling and the web server responsible for maintaining the system’s user interface as shown in figure x.

The web server executes JavaScript logic that enables user interaction, allowing users to adjust settings and receive updates on system status and noise control outcomes. This comprehensive digital infrastructure ensures seamless operation and interaction with the PSANM system, providing users with a powerful tool for managing noise pollution in construction environments.

\vspace{1em} 

\section{Laboratory Experiment Set-up and Testing} \label{sec:lab}
\begin{figure}[t]
\begin{center}
\includegraphics[width=85mm]{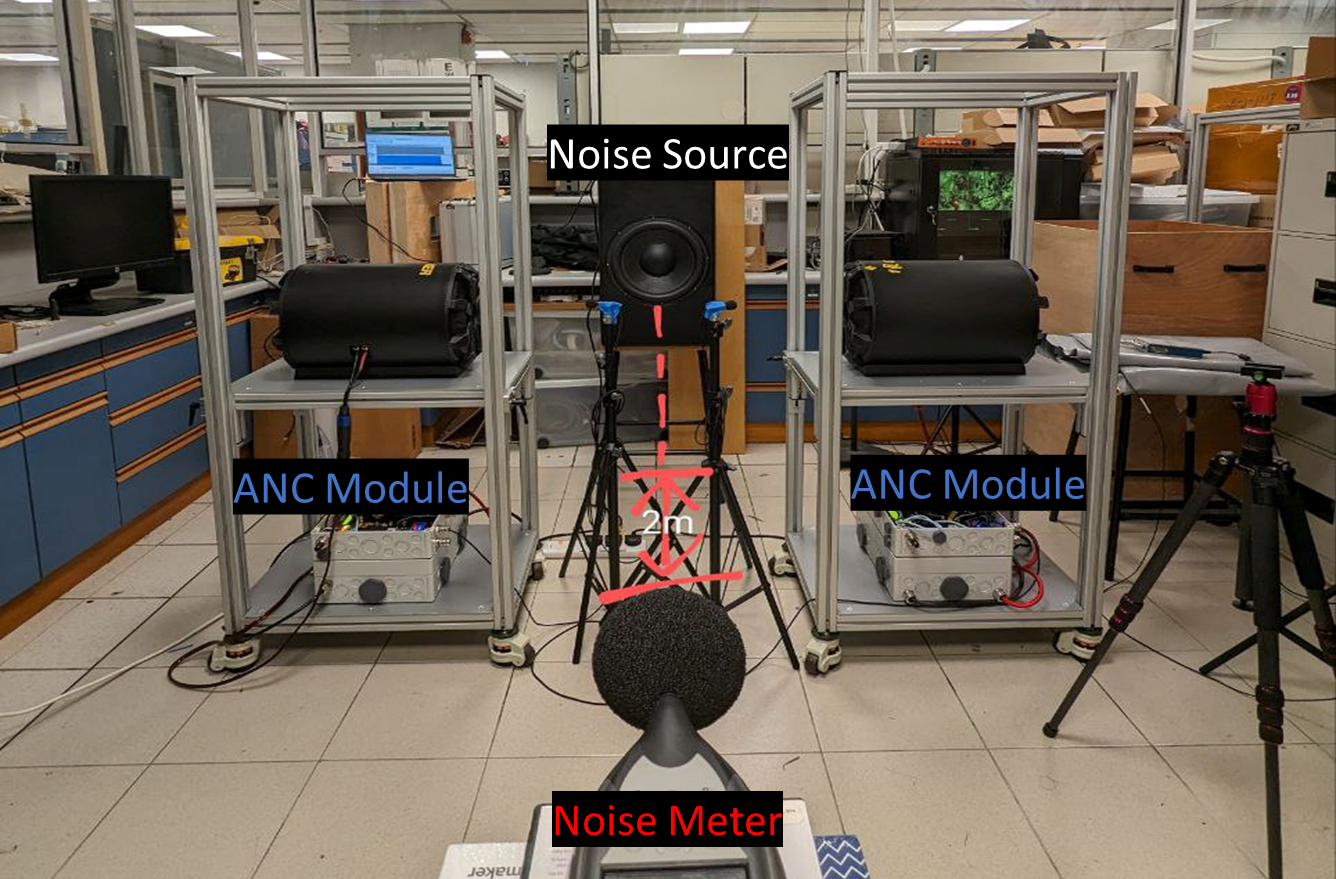}
\end{center}
\label{fig:fig5}
\caption{Laboratory set-up to test and fine-tune the PSANM units}
\vspace*{-3pt}
\end{figure}

\begin{figure}[t]
\begin{center}
\includegraphics[width=85mm]{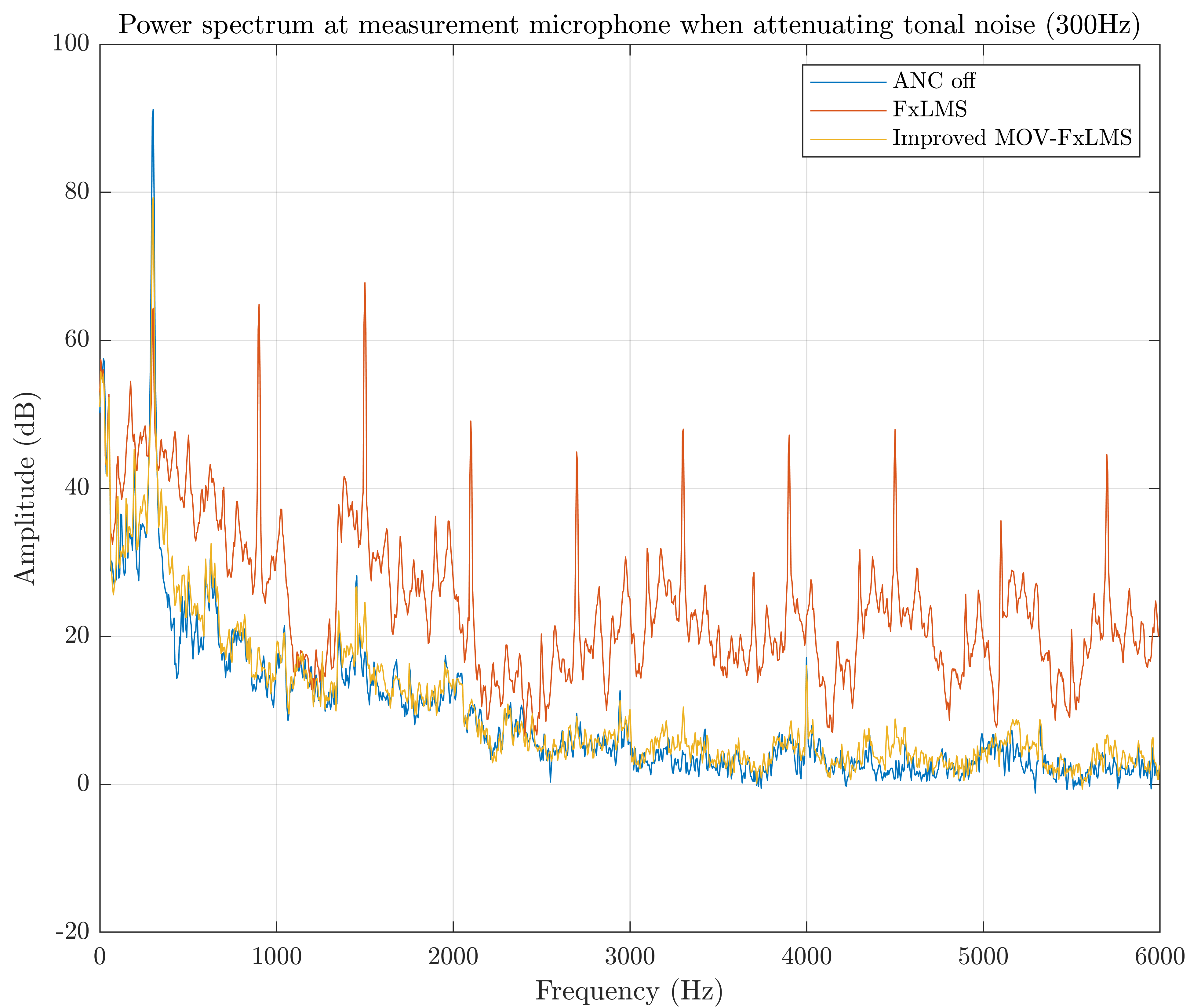}
\end{center}
\label{fig:fig6}
\caption{
    The power spectrum of the measured tonal noise ($150$Hz), when FxLMS and the proposed MOV-FxLMS are tested under output saturation. 
    }
\end{figure}

\begin{figure}[t]
\begin{center}
\includegraphics[width=85mm]{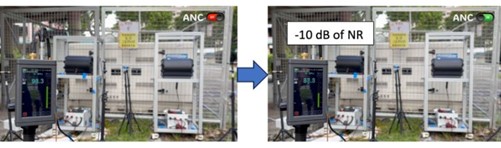}
\end{center}
\label{fig:fig7}
\caption{The actual on-site setup of the PSANM units in front of the Genset noise source where PSANM units achieved a global noise reduction of approximately 10 dB at the frontal working area of the Genset}
\vspace*{-3pt}
\end{figure}

\begin{figure}[t]
\begin{center}
\includegraphics[width=85mm]{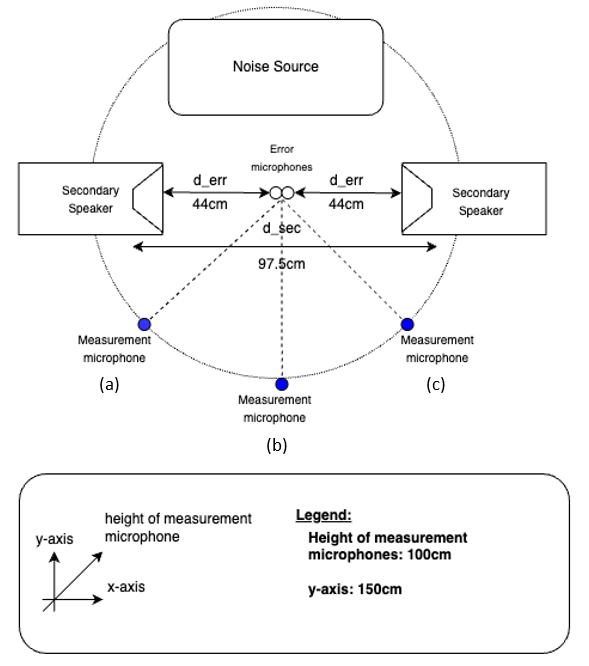}
\end{center}
\label{fig:fig8}
\caption{The position of the PSANM units, error microphones and 3 different monitoring microphone positions (a),(b),(c).}
\vspace*{-3pt}
\end{figure}

\begin{figure}[t]
\begin{center}
\includegraphics[width=85mm]{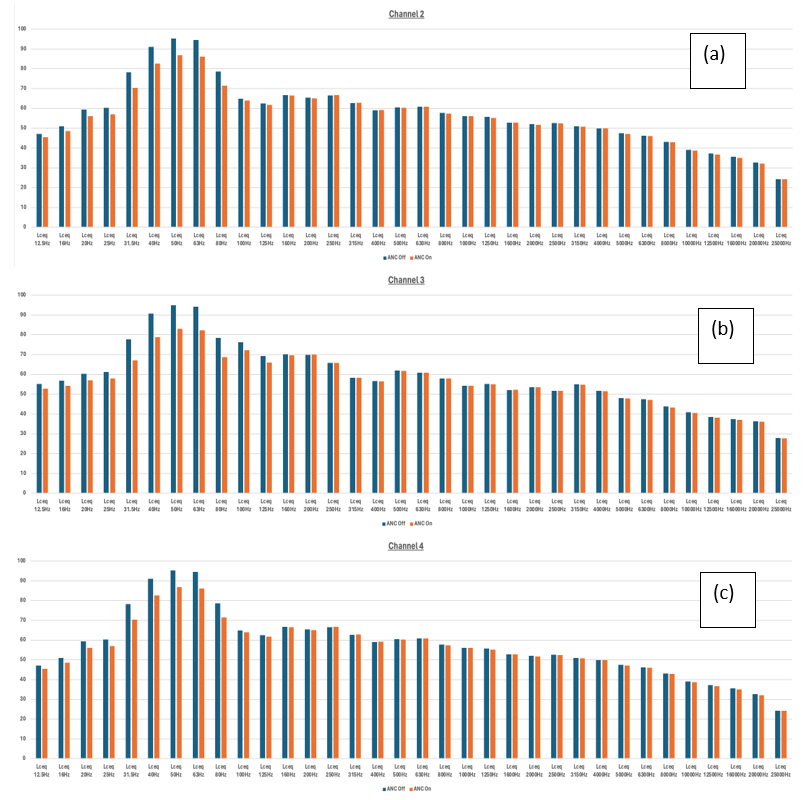}
\end{center}
\label{fig:fig9}
\caption{ The 1/3-octave of noise reduction at 3 different positions (a),(b),(c) show consistent and global noise reduction.}
\vspace*{-3pt}
\end{figure}

To effectively deploy our PSANM system in real-world environments, we recognized the challenges posed by uncontrolled on-site conditions, such as traffic noise and other urban sounds. To address these factors, we conducted a series of controlled laboratory experiments to evaluate the feasibility and noise reduction performance of the PSANM units and to optimize their parameters for maximum efficacy.

In particular, to assess global noise control in free space, the distance \(d\) between the primary noise source and the secondary sources was identified as a critical factor in determining the amount of noise reduction, given a fixed number of secondary sources. Typically, achieving a 10 dB attenuation in total noise power requires a separation normalized distance \((d/\lambda)\) of up to 0.1 for a single secondary source, where \(\lambda\) is the acoustic wavelength\cite{lam2021ten}.

Our laboratory experiments focused on replicating real-world noise conditions in a controlled setting. Figure 5 illustrates the arrangement of two PSANM modules placed facing each other in front of a noise source played back by a subwoofer speaker. Various recorded construction machinery noises and simulated noises were used to evaluate the performance of the PSANM system under different algorithms, parameters, and different noise levels.

In an experiment, a simulated tonal noise at 150 Hz was used as the primary noise, with the constraint set near the peak power of the amplifier. Figure 6 shows the frequency spectrum of the measurement microphone. It can be observed that the conventional FxLMS algorithm generates many harmonic signals at other frequencies due to its lack of output power limitation. This results in the control signal exceeding the output constraint when dealing with large amplitude disturbances, thus causing significant distortions in the anti-noise signal. In contrast, the proposed improved MOV-FxLMS algorithm constrains the output power, effectively reducing the disturbance and avoiding harmonics at other frequencies caused by output saturation.

\section{On-site Genset Noise Control} \label{sec:insitu}

The predominant noise at construction sites is typically stationary and characterized by strong low-frequency components, particularly below 150 Hz, making it an ideal candidate for active noise control applications directly at the source. To test our PSANM system, we identified an operational generator set (Genset) in an open field near a roadside to analyze real noise sources. This on-site deployment aimed to assess the system's performance in a typical construction environment, where noise levels can be unpredictable and influenced by various external factors.

The Genset noise exhibits a high sound pressure level exceeding 90 dBC, often reaching around 100 dBC, with a significant resonance at 77 Hz and slight variations around this frequency. As shown in Fig. 7, we set up two PSANM units in front of the Genset. Operating in feedback mode, the PSANM units achieved a global noise reduction of approximately 10 dB in the frontal area of the Genset. Furthermore, we conducted comprehensive noise reduction measurements at various locations around the Genset, as illustrated in Fig. 8. These measurements consistently demonstrated significant noise reduction, averaging close to 10 dB at low frequencies ranging from 31.9 Hz to 125 Hz. This performance was observed at distances up to 1.5 meters from the Genset and at a height of 1 meter from the ground level, as shown in Fig. 9.

By demonstrating the PSANM system's efficacy in both controlled laboratory settings and on-site machinery noise control, we have laid the groundwork for broader adoption and further investigation into deploying the PSANM in a distributed manner around noise sources. Further minimization of PSANM hardware modules and optimization of the PSANM software architecture and algorithms will facilitate scaling the PSANM units for wide-scale deployment around larger noise sources.

\vspace{1em} 

\section{Conclusions} \label{sec:conclusion}

In conclusion, the development and testing of our PSANM system in both laboratory and on-site deployment demonstrate its effectiveness in managing low-frequency noise emissions from construction machinery. In-situ tests showed a significant global noise reduction of 10 dB, effectively halving the perceived loudness. The use of the MOV-MFxLMS algorithm was crucial for maintaining power output constraints while adapting to sudden noise disturbances. Both feedforward and feedback ANC have been implemented and can be extended to distributed active noise control.

The PSANM system can complement traditional passive noise mitigation approaches, offering broader and more effective noise control coverage. This patent-pending system has the potential for further enhancement and streamlining, enabling the creation of low-cost, mass-producible units for better noise control and broader adoption in urban environments.

\vspace{1em} 
\section*{Acknowledgment}
This research is supported by the Ministry of Education, Singapore, under its Academic Research Fund Tier 2 (MOE-T2EP20221-0014)
\vspace{1em} 

\bibliographystyle{IEEEbib}
\bibliography{mybib}

\end{document}